\documentclass{ceurart}

\usepackage{listings}
\usepackage{todonotes}
\usepackage{algorithm2e}

\copyrightyear{2025}
\copyrightclause{Copyright for this paper by its authors.
  }

\conference{}
\begin{document}

\title{Generating Reliable Adverse event Profiles for Health through Automated Integrated Data (GRAPH-AID): A Semi-Automated Ontology Building Approach}


\author[1]{Srikar Reddy Gadusu}[%
orcid=0009-0008-2606-4746,
email=srikarre@ksu.edu,
]
\cormark[1]
\address[1]{Kansas State University, Manhattan KS 66502, USA}

\author[2]{Larry Callahan}[%
email=lawrence.callahan@fda.hhs.gov,
]
\address[2]{U.S. Food and Drug Administration}

\author[2]{Samir Lababidi}[%
orcid=0000-0002-1406-0448,
email=samlab88@hotmail.com,
]

\author[2]{Arunasri Nishtala}[%
email=arunasri.nishtala@fda.hhs.gov,
]

\author[1]{Sophia Healy}

\author[1]{Hande Küçük McGinty}[%
orcid=0000-0002-9025-5538,
email=hande@ksu.edu,
url=https://www.koncordantlab.com/,
]
\cormark[1]

\cortext[1]{Corresponding author.}

\begin{abstract}
  As data and knowledge expand rapidly, adopting systematic methodologies for ontology generation has become crucial. With the daily increases in data volumes and frequent content changes, the demand for databases to store and retrieve information for the creation of knowledge graphs has become increasingly urgent. The previously established Knowledge Acquisition and Representation Methodology (KNARM) outlines a systematic approach to address these challenges and create knowledge graphs. However, following this methodology highlights the existing challenge of seamlessly integrating Neo4j databases with the Web Ontology Language (OWL). Previous attempts to integrate data from Neo4j into an ontology have been discussed, but these approaches often require an understanding of description logics (DL) syntax, which may not be familiar to many users. Thus, a more accessible method is necessary to bridge this gap. This paper presents a user-friendly approach that utilizes Python and its rdflib library to support ontology development. We showcase our novel approach through a Neo4j database we created by integrating data from the Food and Drug Administration (FDA) Adverse Event Reporting System (FAERS) database. Using this dataset, we developed a Python script that automatically generates the required classes and their axioms, facilitating a smoother integration process. This approach offers a practical solution to the challenges of ontology generation in the context of rapidly growing adverse drug event datasets, supporting improved drug safety monitoring and public health decision-making.
\end{abstract}


\maketitle

\section{Introduction}
In the digital era, the unprecedented growth of data presents both substantial challenges in knowledge management and profound opportunities for discovery. As data accumulates rapidly across various sectors, the imperative to organize, search, and efficiently manage this data is more crucial than ever. Ontologies, fundamental to the Semantic Web, are pivotal in structuring this data, thereby enhancing retrieval capabilities and facilitating knowledge discovery. However, the pace at which data grows and changes demands innovative solutions for ontology creation that can keep up, a task that has traditionally been hampered by complex technological barriers \cite{mbe-odp}, \cite{owl2-primer}, \cite{rdf-spec}.

Traditional ontology development often necessitates a deep understanding of formal logics, such as the Web Ontology Language (OWL), which can be a significant barrier for those without specialized training in description logics (DL) syntax \cite{modont}, \cite{nre-odp}. Although the Knowledge Acquisition and Representation Methodology (KNARM) \cite{mcginty2018knowledge} offers a systematic approach to constructing knowledge graphs, its integration with modern NoSQL databases like Neo4j reveals substantial gaps, particularly in aligning these databases with OWL-based semantic frameworks. Neo4j's capabilities in managing intricate data relationships are robust, yet underutilized in Semantic Web projects due to these integration challenges.

This gap is particularly consequential in the context of high-stakes public health data, such as that from the Food and Drug Administration’s FDA Adverse Event Reporting System (FAERS) \cite{FAERSDB} and Vaccine Adverse Event Reporting System (VAERS) \cite{VAERSDB}. These datasets are critical, as the FAERS dataset contains information on adverse drug events reported during medication use, and the VAERS dataset contains information on symptoms reported after patients receive vaccinations. These datasets provide a rich resource for generating hypotheses and guiding public health decisions. Efficiently integrating this data into semantic frameworks could transform how researchers and policymakers access healthcare data and infer new insights, enhancing responses to public health challenges.

Addressing these issues, this paper proposes a user-friendly approach using Python and the rdflib library to simplify the connection between Neo4j databases and OWL ontologies. Our methodology not only automates the generation of classes and axioms from Neo4j’s data structures but also makes the ontological data more accessible to a broader range of users. By applying this method to the FAERS dataset as a use case, we showcase how our approach supports dynamic ontology development, which is essential for adapting to rapidly changing data landscapes. The enhanced accessibility and practicality of our method allows for the generation of actionable hypotheses from complex datasets, thereby expanding the capabilities of public health research and policy formulation.

\section{Methodology}
To further automate the KNowledge Acquisition and Representation Methodology (KNARM) \cite{mcginty2018knowledge}, three tasks were implemented. First, adverse events data from the Food and Drug Administration’s FDA Adverse Event Reporting System (FAERS) and Vaccine Adverse Event Reporting System (VAERS) were modeled. Second, a graph database was developed to efficiently store and query nested, interconnected data extracted from the FAERS database. Third, an OWL ontology was constructed to represent the semantic relationships embedded within these datasets. The dataset presented specific challenges, which are discussed in the following subsections.

\begin{figure}[h]
\centering
\includegraphics[width=0.95\linewidth]{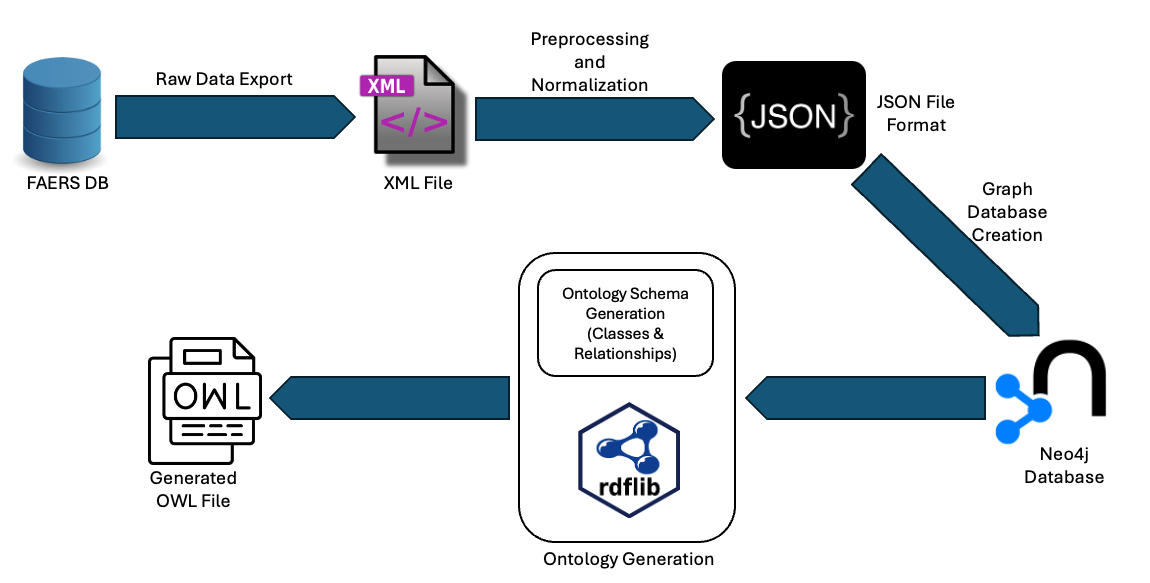}
\caption{Overview of the semi-automated ontology construction workflow. Raw adverse event data is exported from the FDA Adverse Event Reporting System (FAERS) in eXtensible Markup Language (XML) format, then preprocessed and converted into JavaScript Object Notation (JSON). The structured JSON data is loaded into a Neo4j graph database using Cypher queries. Ontology schema generation (classes and relationships) is performed using RDFLib (Resource Description Framework Library for Python), and the resulting ontology is serialized into Web Ontology Language (OWL) format.}
\label{fig:ontology-workflow}
\end{figure}

In addition to FAERS data processing, the VAERS dataset \cite{VAERSDB} was also incorporated into the Neo4j database for visualization purposes. The VAERS data are distributed annually in CSV format, comprising three files for each year: the VAERS Data file (containing patient IDs), the Symptoms file (listing reported symptoms), and the Vaccine file (documenting administered vaccines). Each VAERS ID corresponds to a unique patient record. Cypher queries were implemented to ingest these CSV files and transform them into nodes and relationships within the Neo4j database. The resulting graph structure is illustrated in Figure \ref{VAERS}. This modeling approach closely parallels the FAERS data integration described in the subsequent sections.

\subsection{Modeling the Data and Creating the Database} 
The KNARM methodology and its individual steps have been described in detail in previous work \cite{mcginty2018knowledge}. In this study, attention is focused on the steps of KNARM that were further enhanced through the proposed workflow, specifically: \textit{Meta-Data Creation and Modeling}, \textit{Database Formation}, and \textit{Semi-Automated Ontology Building}. This section describes the modifications introduced to the \textit{Meta-Data Creation and Modeling} and \textit{Database Formation} steps, as illustrated in the overall workflow diagram (Figure~\ref{workflow}).

\subsubsection{Data Collection from FAERS}

The Food and Drug Administration (FDA) maintains the FDA Adverse Event Reporting System (FAERS), a database for collecting and sharing adverse events associated with the use of commercially available drugs. The database consists of quarterly reports beginning in 2004. Each report includes adverse events reported by patients, as well as information on the drugs administered. The dataset provides extensive information that can support hypothesis generation, particularly in identifying drug combinations associated with adverse events, which are often insufficiently studied due to the cost and complexity of individual clinical testing and analysis. Quarterly data from the FAERS database were obtained for this study \cite{FAERSDB}. The acquired data were provided in XML format, containing comprehensive details on adverse events, including safety reports with drug identifications, patient demographics, and reported reactions.

Prior to data processing, a comprehensive review of the database structure was conducted to document all available fields and familiarize with the dataset schema. Several fields utilized numeric codes mapped to standardized vocabulary terms; for example, patient age groups were encoded with values from 1 to 6, representing: 1 = Neonate, 2 = Infant, 3 = Child, 4 = Adolescent, 5 = Adult, and 6 = Elderly.

All relevant fields containing such mappings were identified and documented to ensure accurate interpretation during later stages of data transformation and analysis. Based on this review, a preliminary data model was designed to define the Neo4j graph structure, specifying how nodes and relationships would be created during data ingestion into the graph database.

\subsubsection{Data Conversion} 

Given the structured yet nested nature of the XML data from the FAERS dThe XML data from the FAERS database exhibited a highly nested hierarchical structure. Each parent \texttt{SafetyReport} element contained multiple child elements describing patient demographics, drug consumption, and associated adverse events. Furthermore, child elements often contained additional nested sub-elements. For instance, within each \texttt{SafetyReport}, a \texttt{Patient} element included further child elements such as \texttt{Drug} and \texttt{Reaction}, each of which contained their respective attributes and nested data. Due to this complexity, direct conversion into a flat tabular format (e.g., CSV) posed significant challenges. Flattening the data into rows risked introducing data duplication and loss of relational context, compromising data integrity within the subsequent graph database.

To preserve the nested relationships and minimize data loss, the XML data were first transformed into JSON format. This approach was selected due to JSON’s inherent ability to represent nested hierarchical structures directly, thereby facilitating subsequent processing steps and avoiding the complications that arise with flat formats such as CSV.

During data inspection, inconsistencies across individual safety reports were observed. Certain reports contained complete information, including patient details, administered drugs, and reported adverse events, while others lacked some of these critical elements. The absence of drug names or adverse event terms in particular limited the utility of such reports for downstream analysis, especially when identifying patterns of adverse events potentially associated with drug combinations. Consequently, safety reports lacking either drug or adverse event information were excluded from further processing.

A Python script was developed to automate data cleaning and filtering. The script systematically removed unnecessary elements and excluded safety reports missing either drug or adverse event identifiers. The resulting filtered dataset, still in XML format, was then converted into JSON using a Python-based parser to support subsequent database ingestion and ontology development stages.

\begin{figure}[h]
\centering
\includegraphics[width=0.75\linewidth]{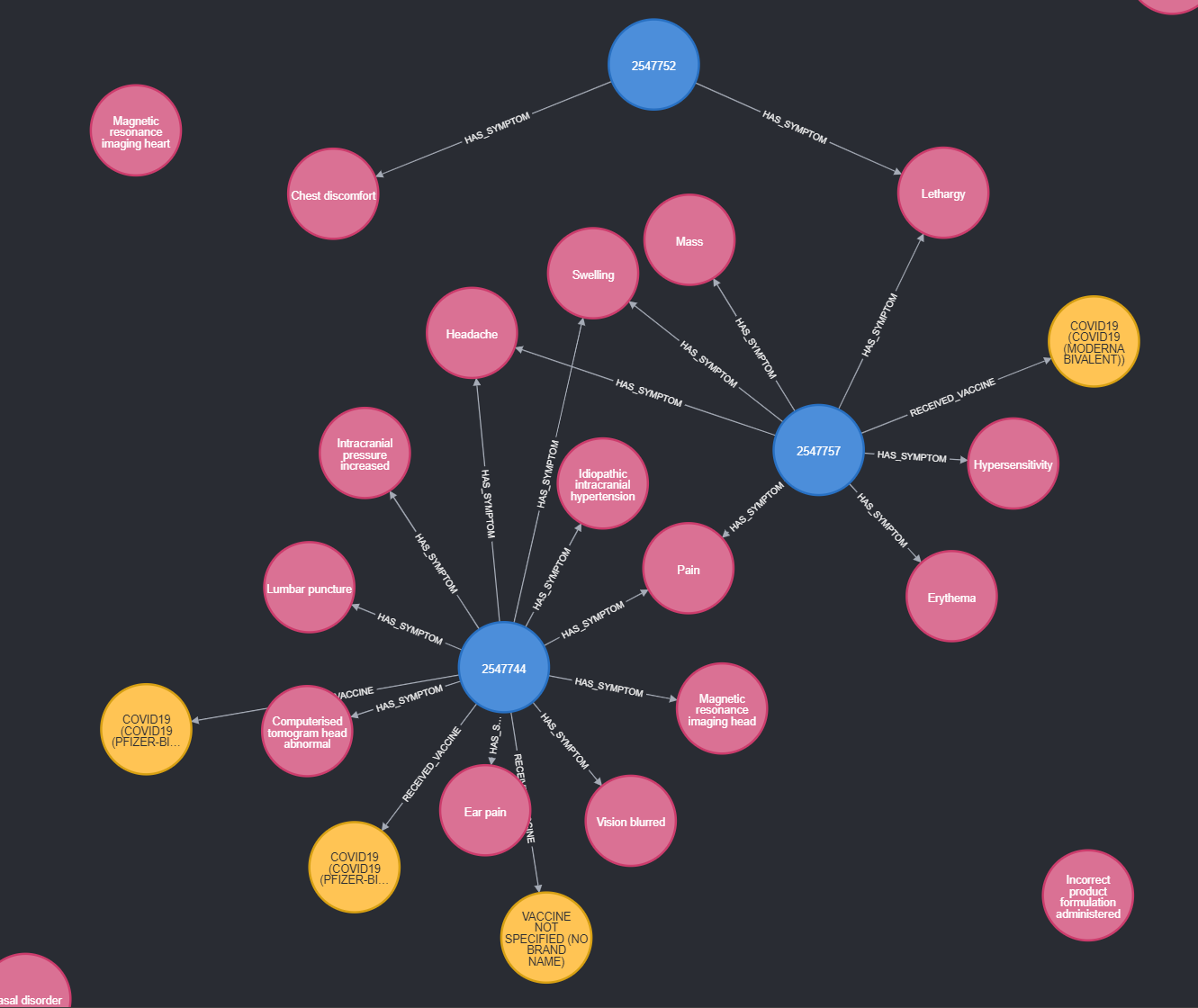}
\caption{Network diagram illustrating two patient nodes (identified by unique IDs) sharing common symptoms within the VAERS dataset.}
\label{VAERS}
\end{figure}

An example of the resulting graph structure from the VAERS dataset is shown in Figure~\ref{VAERS}.

\subsubsection{Neo4j Database Generation} 

The JSON-formatted dataset was ingested into the Neo4j graph database. Prior to data loading, the Awesome Procedures On Cypher (APOC) library was installed to extend the database’s capabilities, including functions for handling JSON files. The ingestion process employed the \texttt{apoc.load.json} function, which reads JSON files directly from the designated project directory.

Data importation proceeded through several structured steps:

\begin{itemize}
\item \textbf{Unwinding the Data:} The \texttt{UNWIND} function was used to iterate over each element within the array of safety reports extracted from the JSON file.
\item \textbf{Node Creation and Matching:} The \texttt{MERGE} function was utilized to create new nodes or match existing ones based on unique identifiers. For each safety report, a corresponding \texttt{SafetyReport} node was established and populated with attributes using the \texttt{SET} command.
\item \textbf{Linking Nodes:} Relationships were established between \texttt{SafetyReport} and \texttt{Patient} nodes. Each patient was linked to associated drugs via the \texttt{TOOK} relationship and to reported adverse events via the \texttt{EXPERIENCED} relationship.
\item \textbf{Property Setting and Relationship Building:} Additional properties were assigned to nodes, and relationships were iteratively constructed using \texttt{FOREACH} loops to process the lists of drugs and adverse events contained within each safety report.
\end{itemize}




\subsection{Creating an OWL file} 

As part of the \textit{Semi-Automated Ontology Building} step, an ontology was constructed to formally represent adverse events recorded in the FAERS dataset, focusing on drug combinations potentially associated with reported outcomes. Due to the extensive size of the full dataset, a representative subset was selected for ontology creation to ensure efficient processing and to allow careful modeling of relevant entities and relationships. The Neo4j database schema was carefully designed to align with the intended ontology structure, facilitating seamless extraction of entities and relationships during the ontology population phase.

Ontology generation was conducted using a custom-developed Python application based on the \texttt{rdflib} library \cite{rdflib}. The process initiated by establishing a connection to the Neo4j graph database through Python’s official Neo4j driver, enabling execution of targeted Cypher queries. These queries were designed to extract essential entities and relationships across four primary conceptual classes: \texttt{SafetyReport}, \texttt{Patient}, \texttt{Drug}, and \texttt{AdverseEvent}. Extracted data included safety report identifiers, patient information, administered drugs, adverse events reported, and the corresponding active substances contained in each drug. The returned data were structured to facilitate efficient translation into Resource Description Framework (RDF) format \cite{rdf-spec}, suitable for ontology construction.

The ontology initialization involved instantiating an RDF graph object using the \texttt{Graph()} function from \texttt{rdflib}. A dedicated namespace was defined to systematically organize Uniform Resource Identifiers (URIs), ensuring consistency, uniqueness, and retrievability of ontology components. This namespace served as the foundational framework for organizing all ontology elements, including classes, properties, and instances.

The ontology schema consisted of formal classes representing the four core entities: \texttt{SafetyReport}, \texttt{Patient}, \texttt{Drug}, and \texttt{AdverseEvent}. Each class was declared in the RDF graph using triples of the form \texttt{(subject, RDF.type, RDFS.Class)}, where each subject was assigned a unique URI from the namespace. These RDF declarations formally defined the ontology’s class hierarchy, enabling semantic reasoning over the represented data.

To capture relationships among entities, object properties were defined including \texttt{has\_patient}, \texttt{took}, \texttt{has\_reported}, and \texttt{is\_partOf\_causing}. These properties were declared as \texttt{OWL.ObjectProperty} instances, while \texttt{has\_activesubstance} was modeled as an \texttt{OWL.DatatypeProperty} to represent literal data values such as active substances within drugs. All properties were assigned unique URIs within the namespace and incorporated into the RDF graph through \texttt{g.add()} operations.

To ensure logical consistency within the ontology, domain and range restrictions were explicitly specified for each property. For example, the property \texttt{has\_patient} was constrained with \texttt{SafetyReport} as its domain and \texttt{Patient} as its range, formalized through triples of the form \texttt{(property, RDFS.domain, Class)} and \texttt{(property, RDFS.range, Class)}. These constraints allowed for well-defined semantic relationships suitable for inference and reasoning tasks.

Following schema definition, the RDF graph was populated with instances corresponding to individual entities retrieved from the Neo4j database. For each unique occurrence of \texttt{SafetyReport}, \texttt{Patient}, \texttt{Drug}, and \texttt{AdverseEvent}, an instance was created and linked to its class using \texttt{RDF.type} statements. Each instance was uniquely identified via its assigned URI. Additional subclass relationships were introduced where applicable, using \texttt{RDFS.subClassOf} statements to capture hierarchical structures or specialized groupings present in the data.

To further enforce semantic integrity and complex domain constraints, OWL restrictions were applied. These were modeled using blank nodes as anonymous class expressions. Each blank node was declared as an \texttt{OWL.Restriction} associated with a specific property via \texttt{OWL.onProperty}. Value constraints were then applied using \texttt{OWL.allValuesFrom} or \texttt{OWL.someValuesFrom} to specify permissible target classes for each property. These restrictions allowed modeling of more nuanced domain semantics, particularly for properties such as \texttt{is\_partOf\_causing}, which capture causal relationships between drugs and adverse events.

Upon completion of schema construction, instance population, and restriction modeling, the RDF graph was serialized into an OWL file using the \texttt{g.serialize()} function. The resulting ontology file, saved in \texttt{.owl} format, could be inspected, validated, and visualized using the Protégé ontology editor \cite{protege}. This semi-automated approach substantially streamlined ontology development and enabled formal semantic representation of complex healthcare data.






\section{Conclusion and Future Work}
This study established a semi-automated ontology generation framework integrating data from the Food and Drug Administration’s Adverse Event Reporting System (FAERS) and Vaccine Adverse Event Reporting System (VAERS), using Neo4j for graph-based data storage, combined with the Web Ontology Language (OWL) for semantic modeling, implemented via the Python \texttt{rdflib} library.

The presented approach addresses key challenges associated with incorporating large-scale, dynamically evolving public health datasets into formal semantic frameworks. By automating data acquisition, transformation, and ontology generation, the system enhances the efficiency and reproducibility of knowledge graph construction while preserving semantic richness. This enables downstream applications such as hypothesis development, risk signal detection, and integrative analysis of drug safety data.

By streamlining ontology development, the proposed system facilitates broader access to complex biomedical datasets by both public health researchers and regulatory agencies. For example, the ontology could support hypothesis generation concerning emerging adverse drug events caused by multi-drug interactions, or assist in identifying patient subgroups at higher risk for specific adverse reactions. Such capabilities have the potential to accelerate pharmacovigilance efforts, improve public health responses, and ultimately support patient safety.

Future work will focus on further automating the ontology generation pipeline to enhance scalability and reduce manual intervention. Planned enhancements include developing direct ontology class generation from Neo4j data and enabling bi-directional conversion between OWL ontologies and Neo4j databases. These capabilities would broaden the applicability of this methodology across diverse domains requiring rapid, accurate semantic modeling for knowledge engineering and decision support.


\appendix

\section{Online Resources}
The source code is available in the below github link 
\begin{itemize}
\item \href{https://github.com/koncordantlab/Neo4jToOWL.git}{https://github.com/koncordantlab/Neo4jToOWL.git}
\end{itemize}

\end{document}